\documentclass[10pt,journal,compsoc]{IEEEtran}

\ifCLASSOPTIONcompsoc
  \usepackage[nocompress]{cite}
\else
  \usepackage{cite}
\fi

\usepackage{epsfig}
\usepackage{graphicx}
\usepackage{amsmath}
\usepackage{amssymb}
\usepackage{dirtree}
\usepackage{subcaption}
\usepackage{booktabs}
\usepackage{amsthm}		
\usepackage[lined,boxed]{algorithm2e}
\usepackage{paracol}
\globalcounter{algocf}
\SetAlFnt{\small}

\usepackage{setspace}		
\usepackage{siunitx}     
\usepackage[bottom]{footmisc}	
\usepackage{mathtools}
\usepackage{multirow}

\newcommand{\PD}{\mbox{PD}}
\newcommand{\DI}{\mbox{DI}}

\usepackage{array}

\newcommand{\Vspcx}{\vphantom{\Huge x}}
\newcommand{\VspcX}{\vphantom{\Huge X}}

\graphicspath{benchmarks/}

\begin{document}

\title{The Cascading Metric Tree}
\author{{Jeffrey~Uhlmann,~ Miguel~R.\ Zuniga}%
        
\IEEEcompsocitemizethanks{\IEEEcompsocthanksitem Prof.\ Uhlmann is a faculty of the Department
of Electrical Engineering and Computer Science, University of Missouri, Columbia, MO, 65211.\protect\\
\IEEEcompsocthanksitem M.\ Zuniga is head of biosystems analytics at New Generation Software.}
\thanks{Manuscript received September 1, 2021; revised December 26, 2021.}}

\markboth{Journal of \LaTeX\ Class Files,~Vol.~14, No.~8, August~2015}%
{The Cascading Metric Tree}

\IEEEtitleabstractindextext{%
\begin{abstract}
This paper presents the Cascaded Metric Tree (CMT) for efficient satisfaction of metric search queries over a dataset of $N$ objects. It provides extra information that permits query algorithms to exploit all distance calculations performed along each path in the tree for pruning purposes. In addition to improving standard metric range (ball) query algorithms, we present a new algorithm for exploiting the CMT cascaded information to achieve near-optimal performance for $k$-nearest neighbor (kNN) queries. We demonstrate the performance advantage of CMT over classical metric search structures on synthetic datasets of up to 10 million objects and on the $564K$ Swiss-Prot protein sequence dataset containing over $200$ million amino acids. As a supplement to the paper, we provide reference implementations of the empirically-examined algorithms to encourage improvements and further applications of CMT to practical scientific and engineering problems. 
\end{abstract}

}

\maketitle
\IEEEdisplaynontitleabstractindextext

%
\IEEEpeerreviewmaketitle

\section{Introduction}\label{sec:introduction}

\IEEEPARstart{I}{n} this paper we examine new methods, as well as combinations
of known methods, for augmenting metric search structures to provide a level of 
performance necessary for practical applications, e.g., involving large-scale protein
and genomic sequence datasets. Our first contribution is the use of {\em metric 
cascading}, in the form of a {\em cascaded metric tree} (CMT) to permit every 
distance calculation performed along a path in a tree to provide an additional 
pruning test at every subsequent node along that path. In other words, if $P$ 
distance calculations have been performed along the path from the root node to 
a given node, then $O(P)$ fast independent scalar-inequality tests (i.e., not requiring 
computationally-expensive metric distance calculations) are available for pruning
the search at that node.

We formally define a general family of cascaded metric tree structures for which 
the number of distance calculations per node is parameterized. However, our goal
in this paper is to define a search structure and associated query algorithms that 
maximally exploit information from each distance calculation such that near-optimal 
performance can be expected without need for application-specific tuning. To this 
end, our tests only examine the most basic form with no parametric tuning 
of any kind.

\section{Background}\label{sec:background}

\IEEEPARstart{P}{erhaps} the simplest generalization of the familiar 1d (scalar) 
balanced binary search tree (BST) to accommodate general metric search queries 
can be constructed by choosing one of the $n$ objects in a dataset and computing 
the distance (radius) from it to the remaining objects, with the object and the median 
distance stored at the root node. Applying this process recursively such that objects 
less than the median radius at a given node form the left subtree, and the remainder 
form the right subtree, yields a balanced {\em metric} binary search tree (MBST) such 
that the path to a given object is uniquely determined by a sequence of distance 
calculations starting at the root node. For example, if the distance between a given 
query object $s$ and the object $r$ stored at the root node is less than the radius 
stored at the node then $s$ must be in the left subtree, otherwise it must be in the 
right subtree. This provides a recursive search algorithm that is guaranteed to find 
$s$, or determine that object $s$ is not in the tree, using only $O(\log(N))$ distance 
calculations

Generalization of the MBST search algorithm to satisfy queries asking for all objects
within a given distance from a query object, referred to as a metric {\em range} or
{\em ball} query, are directly analogous to the satisfying of interval range queries in 
a BST where both subtrees of a node must be searched if the query range spans
the partitioning value. In the case of the simple MBST, a metric range query can only
prune a subtree from the search process when the metric ball determined by the 
object and radius at a given node does not intersect the query volume (ball). This 
requires only the calculation of the distance between the query object and the object 
at the node, followed by a triangle inequality test involving their respective radii. Thus, 
the triangle inequality property of metric distances is absolutely necessary for 
correctness of the algorithm when pruning is applied\footnote{The books of
Samet \cite{sametbook} and Zuzula {\em et al.} \cite{zezula2006similarity}, 
and the review article by Chavez {\em et al.}  \cite{chavez2001searching},
provide outstanding coverage of the literature on metric search structures
and algorithms.} 

As described thus far, the simple MBST only permits left subtrees to be pruned because 
there is no available bound on the spatial extent of objects in the right subtree. If in 
addition to the median, however, information defining the minimum-volume 
{\em bounding ball} is also stored at each node then an additional criterion for pruning 
will be available at each node during the search process. Unfortunately, this bounding 
ball cannot necessarily be made tight except in the case of vector spaces (which permit
arithmetic operations to be performed on objects, e.g., to create a new object as the
mean of two other objects) because there is no way in general to construct a virtual 
``spatial centroid'' object in the general case of black-box distance functions and objects. 
However, it {\em is} possible to identify an object for which the maximum radius to any 
other object is minimized, but this incurs the cost of a potentially quadratic number of 
distance calculations during the tree construction process. 

Clearly there are many degrees of freedom available during the tree construction 
process that can be tailored to potentially improve MBST query performance, (i.e., 
to reduce the average number of distance calculations per query) based on the 
characteristics of datasets that are deemed typical for a particular application. For 
example, the strict binary structure can be relaxed to a $b$-ary tree with branching-value 
$b$ determined empirically based on tests over representative datasets for a given 
application. Unfortunately, too many degrees of freedom can mask the fact that the 
search algorithm is suboptimal in its sensitivity to application-specific variables. Ideally, 
there would exist a data structure that provides consistent near-optimal performance 
such that minimal improvement can be expected from additional heuristic 
application-specific tuning. In other words, search structures and/or algorithms with 
few or no tunable parameters are likely to exhibit more robust practical performance 
than those for which good performance demands extensive tuning of many parameters.

\section{Metric Search}\label{sec:metric}

\IEEEPARstart{A}{} function $d(x,y)$ defines a metric distance on the set $S$ if for 
elements $x,y,z$ of $S$ if the properties in Table \ref{table:metric} hold.
\begin{table}[htb]
\centering
\caption{Metric Properties $\forall\, x,y,z \in S$}\vspace{-8pt}\label{table:metric}
\begin{tabular}{ l | l }
	\hline
	$d(x,y) \ge 0$ & non-negativity\Vspcx \\
	$d(x,y) = d(y,x)$ & symmetry \\
	$x = y \Rightarrow d(x,y) = 0$ & identity\\
	$d(x,z) \leq d(x,y) + d(y,z)$ & triangle inequality \\
	\hline 
\end{tabular}
\end{table}

Given a query object $q$, a search radius $r$, and a set $S$ of objects, a metric range 
query $R(q,r)$ over $S$ reports all objects in $S$ within distance $r$ of the object $q$.  
An exhaustive (brute force) search can satisfy the query by performing $|S|$ distance 
functions, but this may not be practical when $S$ is very large or when the relevant 
distance function $d(x,y)$ is computationally expensive to evaluate. However, the metric 
triangle inequality property offers a means for potentially reducing the number of distance 
calculations needed to satisfy such queries. For example, consider object $p \in S$ and 
$C \equiv \{ S \setminus p \}$.  Also assume that the nearest and farthest objects in $C$ 
from $p$ are at known distances  $d(p,p_n)$ and $d(p,p_f)$, respectively. Given these 
conditions there exist queries $R(q,r)$ for which the performing of only one additional 
distance calculation $d(p,q)$ establishes that {\em none} of the objects in $C$ satisfy 
the query. For example, in the case when the query object $q$ is ``far'' from $p$ in the 
sense that it is farther from $p$ than any object in $S$, i.e, $d(q,p) > d(p,p_f)$, then 
$d(q, p) > d(p,p_f) + r$ implies $d(q,x) > r,  \: \forall \, x  \, \in \{C\}$:
\begin{quote}
\begin{proof}
  Let $ d(q, p) - d(p,p_f) > r$,  so
  \begin{flalign*}
    &d(q, x)\! +\! d(x, p)\! -\! d(p, p_f) > r  ~\text{({triangle inequality})}\\
    &d(q, x) > r + [d(p,p_f) - d(x, p)] \\
    &d(q, x) > r  ~~~~ \text{({since $d(p,p_f)\! -\! d(x, p)$ is positive})} \\
     \qedhere     
  \end{flalign*}
\end{proof}
\end{quote}
An analogous result holds for the case of a ``near'' query object, i.e., $d(q,p) < d(p,p_n)$, 
when $d(q,p) > d(p, p_n) - r$. It is convenient to view $d(p,p_n)$ and $d(p,p_f)$ as 
defining a distance interval $\DI(p, C) \equiv [d(p,p_n), d(p,p_f)]$ that bounds the distances 
from $p$ to the objects in $C$ and which may permit an immediate determination (via the 
triangle inequality) that the query ball {\em cannot} contain any objects in $C$. In other 
words, either of two simple inequality tests may establish that further searching is 
unnecessary. Various such inequalities have been examined in the literature and provide 
the standard means for exploiting metric assumptions for pruning searches of objects in 
large datasets.

The limitation of conventional metric search trees during the query process is that information 
obtained from the computation of a distance calculation between the query object and an object 
stored at a given node is only exploited for pruning purposes {\em at that node}. In the following 
section we describe a means for augmenting the search tree with additional information so that 
the pruning step at the current node can exploit all previous distance calculations performed 
along the path from the root to the current node.

\subsection {Cascaded Metric Trees}

\IEEEPARstart{C}{onsider} 
a dataset that is recursively decomposed into a metric search tree with one or more objects 
stored at each node. Purely for convenience of exposition we will assume a balanced binary tree 
with a single object stored at each node. We refer to the structure as a {\em metric} search 
structure because it is further assumed that any search algorithm must traverse the tree based 
on information obtained from black-box metric distance calculations involving objects stored along the 
sequence of visited nodes. Ideally, this information will permit the search space (i.e., the 
amount of the tree that must be traversed) to be significantly reduced so that the total number 
of performed distance calculations does not greatly exceed the number of objects that are found 
to satisfy the query. 

\begin{figure*}
\centering
\includegraphics[width=0.75\linewidth]{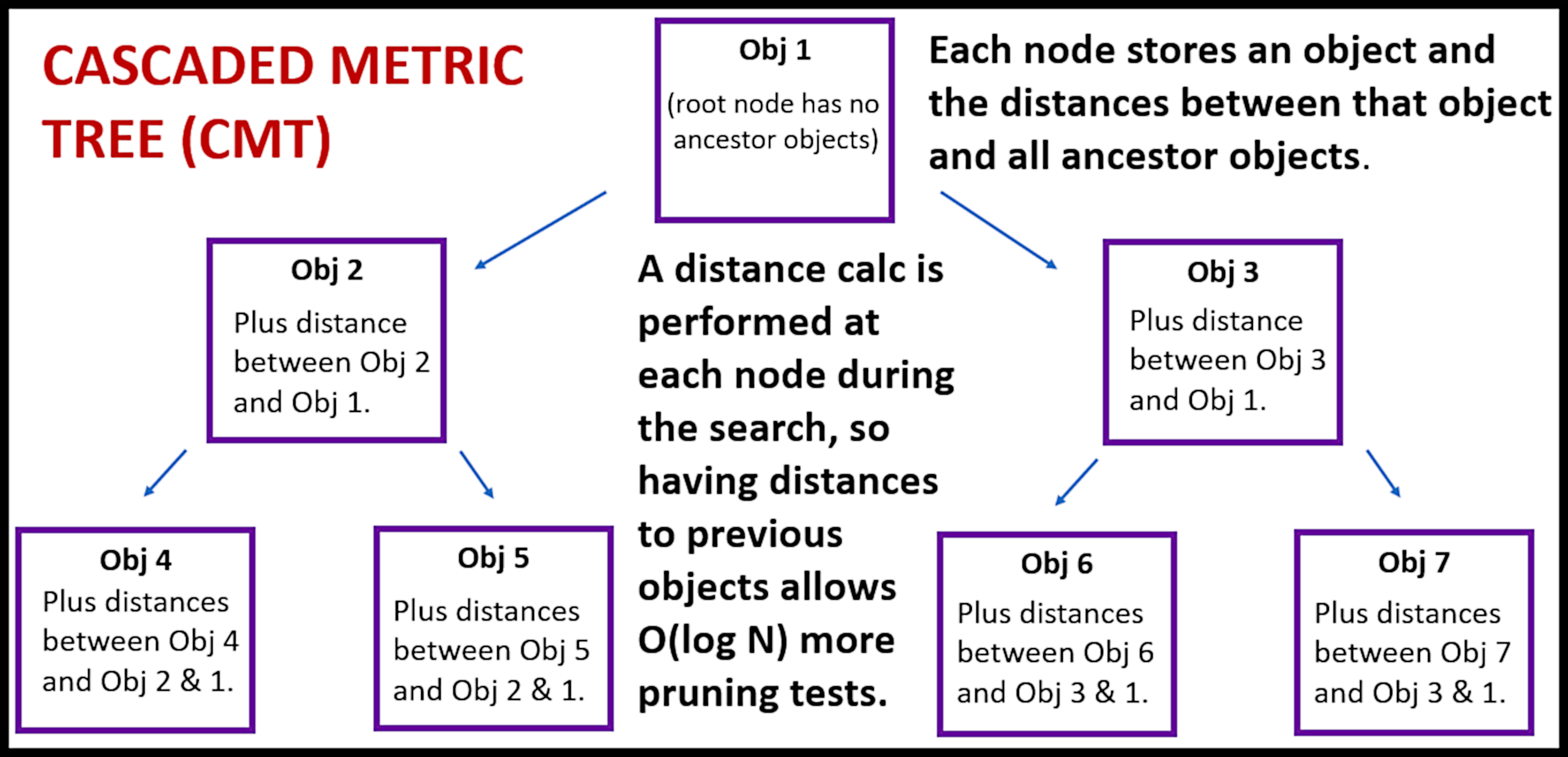}
\caption{In addition to an object, each CMT node contains distances between its object and all ancestor objects. (These distances are pre-computed during construction of the tree.) Because the query object has already computed distances between itself and the ancestor objects of the current node at depth $d$, an extra $d$ triangle-inequality tests can be performed without need for any extra distance calculations.}
\label{fig:tree}
\end{figure*}

The principal feature of the cascaded metric tree (CMT) is that each node contains not only
an object, but also stores the distance between that object and each object along the path
from that node to the root of the tree (see Fig.\,\ref{fig:tree}). Because a distance calculation
must have been performed between the query object and the {\em ancestor} objects visited
along the path to the current node, all the information necessary to perform triangle-inequality pruning tests 
with respect those ancestor objects is available with no need for additional distance calculations. 
In other words, the number of ``free'' pruning tests at each node increases linearly with depth.
More fundamentally, each distance calculation performed at a node during the search process
is exploited in a pruning test at each subsequent node along that path in the tree, as opposed
to a single pruning test at the current node. 

Importantly, the CMT construction algorithm performs only $O(N\log N)$ total distance calculations,
which is the same as that required for conventional linear-space metric search trees, 
despite the fact that CMT stores $O(\log N)$ ancestral distances at each node. This is possible
because the CMT construction process is able to exploit cascaded distance information in a
manner that is analogous to the search process.

To formalize the specification of the tree and search algorithms, we define 
the following notation for information defined for or maintained by the query algorithm:
\begin{itemize}
\item $q$ is the query object.
\item $r\geq 0$ is the query radius. A value of $\infty$ may be implicitly or explicitly defined for 
         the case in which no bound is placed on the size of the query radius. 
\item $k\geq 1$ is an integer limit on the size of the returned query set of nearest objects. 
         A value of $\infty$ (or sufficiently large number greater than or equal to the size of the 
         dataset) may be implicitly or explicitly defined for the case in which no limit is placed on 
         the number of objects that may be returned. 
\item {\em depth} is the number of levels the current node is from the root. Thus, 
         {\em depth}\,-\,1 is the depth of the parent node, {\em depth}~-~2 is the grandparent 
         node, and the ancestor at level {\em depth} is the root node of the overall tree. 
         The value of {\em depth} is dynamically maintained during the search/traversal process 
         and thus is not a query parameter. 
\item {\em dist}[j] is the distance from $q$ to the $j$th ancestor of the current node 
         $1 \leq j <$\;{\em depth}. This indexed list/vector is dynamically maintained during the 
         search/traversal process and thus is not a query parameter. 
\end{itemize}
And we define the following notation for information maintained at each CMT node:
\begin{itemize}
\item {\em near}[i] and {\em far}[i] provide near and far distance interval information with 
         respect to the $i$th ancestor of the node. For example, {\em near}[1] is the distance of the 
         object in the subtree rooted at the current node (which includes $p$) that is nearest to the 
         node-object in the parent, and {\em far}[1] is analogously defined with respect to the farthest 
         object in the subtree from the parent object. More generally, distance interval information with 
         respect to the $i$th ancestor of the current node is similarly available for each index 
         $1 \leq i <$\;{\em depth}. The values {\em near}[0] and {\em far}[0] provide the near and 
         far distances to objects in the subtree beneath the current node, i.e., excluding the object $p$ 
         at the current node. 
\item {\em count} provides the number of objects in the subtree of the current node. This is necessary 
         to satisfy {\em counting} queries, which ask for the number of objects that satisfy a given query 
         without actually retrieving those objects. More specifically, if the query volume completely contains 
         the bounding volume of objects in the subtree of a given node (which can be determined from $p$
         and {\em far}[0]), then the number of satisfying objects is immediately given by {\em count} 
         without need to traverse the subtree. This permits counting queries to be satisfied with complexity 
         sublinear in the number of objects that satisfy the query, e.g., $O(1)$ if the query volume is 
         sufficiently large to enclose all objects in the tree.  
\item {\em left} and {\em right} are the left and right child nodes (subtrees) of the current node.
\end{itemize}

If the query interval/ball fails to intersect the distance interval/annulus for any ancestor $i$ then the 
entire subtree (including the node-object $p$) can be pruned from further consideration. This provides 
$O(\log N)$ separate pruning tests, each of which may immediately terminate search of the current 
subtree. More specifically, the number of pruning tests increases from $O(1)$ at the root node to 
$O(\mbox{\em height})$ at leaf nodes, which is $O(\log N)$ if the tree is balanced.

\begin{table}[htb]
\centering
\caption{Information Relating to the Current Query}\vspace{-8pt}\label{table:queryparms}
\begin{tabular}{c|l}
\hline 
$q$\Vspcx & The query object: goal is to find the $k$-nearest\\ 
~ & objects within distance $r$ of $q$.\\
$r$ & The maximum allowed radius (distance) from the\\
~ & query object $q$. \\
$k$ & The maximum allowed number of objects nearest\\
~ & to $q$ to be returned. \\
~ & ~~~({\em Below is information maintained during}\\ 
~ & ~~~~{\em execution of each query}) \\
{\em depth} & The number of levels the current node is\\ 
~ & from the root. \\
{\em dist}[j] & The distance from $q$ to the node-object\\ 
~ & of the $j$th ancestor of the current node. \\
\hline
\end{tabular}
\end{table}

\begin{table}[htb]
\centering
\caption{Information Stored at Each Node of the CMT}\vspace{-8pt}\label{table:treeparms}
\begin{tabular}{c|l}
\hline
$p$\Vspcx & The node-object, i.e., the object stored at\\
~ & \,the node. \\
{\em near}[i],\Vspcx & The min and max distances from the $i$th\\
{\em far}[i]  & ancestor object to the objects in the subtree.\\
~ & ({\em Index $0$ refers to min and max distances from}\\
~ & \,{\em $p$ to objects in its subtree, excluding $p$ itself}).\\
{\em count}\Vspcx & The number of objects in the subtree.\\
{\em left, right}\Vspcx  & The left (right) child/subtree. \\
\hline
\end{tabular}
\end{table}

It should be recognized that a conventional metric range/ball query (i.e., asking for all objects in the 
tree that are within a specified radius $r$ of a given query object $q$) can be parameterized with $k$ 
implicitly equal to infinity. Similarly, a $k$-nearest query can be parameterized with $r$ implicitly equal 
to infinity. More generally, nontrivial values can be specified for both $r$ and $k$. This allows the use 
of $r$ so that the return set from a $k$-nearest query does not include elements that are of an 
impractically-large distance $r$ from $q$. Similarly, $k$ can be specified to ensure that an 
impractically-large number of objects within distance $r$ of $q$ are returned by a ball query. In practice, 
the efficiency of a $k$-nearest query can often be improved if a nontrivial upper-bound radius $r$ can be 
provided to guide the search. The benefits of this {\em range-bounded} near-neighbor search has long been
recognized \cite{uhlmannImp1991,uhlmannDAconf91,10.5555/338219.338273}.
We further propose additional parameters $nv$ and $dc$ to limit the total 
number of nodes visited and the total number of distance calculations performed, respectively, during the 
query. These additional parameters can be used impose rigid limits on the query complexity and thus 
would provide only approximate query solutions based on a limited search of the tree. However, we will 
not exploit the $nv$ and $dc$ parameters here because our focus in this paper is on {\em exact} metric 
query satisfaction.

More generally, the node-object $p$ in each node can be replaced with a list of $m$ node-objects $p[j]$,  
$1<j\leq m$. Thus each $p[j]$ provides a distinct {\em near/far} distance interval for each descendant, 
which means that two indices are required: {\em near}$[i][j]$ and {\em far}$[i][j]$.  Note that even in 
the case that $m$ is chosen to be the same value $c$ for every node, or for every node at a given level 
of the tree, an explicit integer is required for each node to accommodate the case of a number of objects 
not equal to $c$ at leaf nodes. Note also that in addition to maintaining ancestor distance information with 
respect to $q$, the number of node-objects at each ancestor must also be maintained. 

\begin{table}[htb]
\centering
\caption{Information Stored at Each Node of General CMT}\vspace{-8pt}\label{table:gcmtparms}
\begin{tabular}{c|l}
\hline
$m$\Vspcx & Strictly positive integer giving the number\\
~ & of node-objects at the node.\\
$p[j]$\Vspcx & The list of $m$ node-objects, $1\leq j \leq m$.\\
{\em near}$[i][j]$,\VspcX & Min and max distances from the $i$th ancestor\\
{\em far}$[i][j]$  & object to the objects in the subtree.\\
{\em count}\Vspcx & The number of objects in the subtree.\\
{\em left, right}\Vspcx  & The left (right) child/subtree. \\
\hline
\end{tabular}
\end{table}

\noindent The generalization to $m>1$ objects per node provides a complex tradeoff involving more 
space and more distance calculations per node in return for a multiplicative increase in pruning tests per 
node\footnote{In fact, the number of distance calculations per node can be varied, e.g., more at nodes
near the top of the tree and decreasing to zero at leaf nodes}. At the opposite limit, a leaf-oriented CMT 
(or LCMT) can be defined with {\em zero} objects stored at each interior node, thus offering a means
to reduce the overall number of distance calculations if certain assumptions are satisfied. While these 
generalizations offer vast opportunities for application-specific tailoring, our stated goal for this paper 
is to focus on core performance without any discretionary variables. Therefore, we will assume $m=1$
and leave the examination of alternatives for future work. 

Regarding test results, our principal measure of performance will be the total number of distance 
calculations performed because those calculations are fundamental to the problem and are expected 
to dominate the overall running time for most nontrivial metrics. Although our tests will also include 
information about the number of nodes visited, we note that this is an unreliable measure that can 
easily be ``gamed'' simply by front-loading simple termination tests that cannot actually reduce the 
number of distance calculations -- {\em and may even increase the overall computational overhead} -- 
but can reduce node visitations at the frontier of the query traversal by as much as a factor of two. In
other words, the overall computation cost may be increased while ``{\em number of nodes visited}'' as  
a performance metric spuriously suggests otherwise. (We note that it may be worthwhile to reduce node 
visitations in some contexts, e.g., to reduce external memory accesses, but that choice is available for 
any tree-type data structure. In many respects it is analogous to increasing the branching factor
of a tree to compress its height in order to optimize an application-specific utility function.)  

To summarize, the focus of this paper is the method of {\em metric cascading}, which involves the 
storing of ancillary information at each node to permit distance calculations performed at ancestor 
nodes during a search to subsequently be exploited for potential pruning of the search. In the most 
general formulation this information grows linearly with the depth of each node in the cascading 
metric tree (CMT). Thus, under our assumptions with $N=|S|$, the total space of the CMT increases 
from $O(N)$ to $O(N\log(N))$. This extra information consists of the {\em near} and {\em far} 
distances to objects in the tree rooted at each node with respect to each of the objects stored in the 
{\em ancestor} nodes on the path from the root to the node. This means that every distance calculation 
performed by the search algorithm prior to the current node can be exploited for pruning purposes at 
the current node. This represents a significant departure from traditional metric search structures which 
do not exploit any information from distance calculations performed at previously-visited nodes. This
will be demonstrated respectively for range and $k$-nearest neighbor (kNN) in the next two 
sections.

\section{CMT Range (Ball) Queries}

\IEEEPARstart{T}{he} basic algorithm for satisfying a 
range query on a metric search tree consists of determining
whether the query ball intersects the bounding ball associated
with the currently-visited node. If so then the subtree of that
node must be searched, otherwise it can be eliminated
(pruned) from further search. 

In a conventional search tree, the intersection test is a simple 
application of the triangle inequality based on the distance of the 
query object to the node object, the query radius $r$, and the
bounding radius for the node. What the CMT provides is the
distances between the current node object and the objects
in all of its ancestral nodes. In its traversal to the current node,
the search process has performed distance calculations between 
the query object and those ancestral objects. If the results of
those distance calculations have been accumulated, then each
can be exploited to provide an independent triangle-inequality
pruning test. 

The details of the basic CMT range query algorithm are given in
Appendix\,\ref{app:BasicRQ}. The adjective {\em basic} is applied
because it does not fully exploit all available opportunities for pruning.
Specifically, the stored bounding radius at each node defines a metric ball
within which all objects in the subtree of the node are contained. The
pruning test of the simple range query algorithm checks whether the
query volume intersects the bounding volume of the node and prunes the
search if it does not. However, if the query volume {\em encloses} that
bounding volume, then it can be concluded that all objects in the subtree
must satisfy the query. This means that a {\em collection/aggregation}
operation can be performed to add all of the objects in the subtree
to the retrieval set without need for any distance calculations with the
query object. 

Details of the full CMT range query algorithm with collection are
provided in Appendix\,\ref{app:collectRQ}. 
An extreme example of the benefit provided by collection is the
case in which the search algorithm can determine at the root node that
all objects in the tree are contained within the query volume. In this
case a single distance calculation is sufficient to identify that all
objects in the tree satisfy the query. Collection is particularly
effective when the object stored at each CMT node is selected to be a
radius-minimizing centroid of the objects in its subtree\footnote{The 
CMT construction algorithm used for our tests (Appendix\,\ref{app:bom})
randomly selects the object stored at each node from the set of objects
comprising its subtree, whereas a radius-minimizing centroid would clearly
be a superior choice to maximally exploit collection. However, our test
results demonstrate it to be effective because it is an element of the set 
of objects in the subtree. If the object were not a member of the set 
then the bounding radius required to contain the set would likely be 
impractically large to effectively support collection.}. It can be 
expected that collection will most likely be triggered at nodes lower in 
the tree that have relatively smaller bounding distances, or for queries 
with relatively large search radii.

\subsection{Collection/Aggregation for Counting Queries}

In addition to its use for more efficiently satisfying range queries, 
collection can also be applied to calculate only the {\em size} of 
the return set for a given query without actually performing
retrieval of the objects. Specifically, if each node stores an integer giving
the size of its associated subtree, then a {\em counting query} can be
performed in which only the {\em number} of objects satisfying the
query is computed without returning that set of objects. For example,
instead of performing a collection operation at a given node, the
number of objects in the subtree is simply added to the count without
need to even traverse the subtree.
Counting queries are important in the context of interactive data
analysis applications to permit analysts to refine their queries to
guarantee that subsequent retrieval queries have focused return sets
of manageable size. To be effective, however, results of all 
distance calculations must be maintained in a hash table (or other
kind of efficient map \cite{Cormen2001introduction})
 to avoid redundant calculations during subsequent
queries within the interactive session for a given query object.  

Collection-enhanced query satisfaction also offers potential benefit
in applications for which the distance function used for queries is
actually a surrogate for a far more complex metric or in place of a
measure of similarity for which a tight equivalent distance function
cannot be found. This is common in domains involving protein and
biosequence objects for which the most explicit and intuitively
understood models for assessing {\em similarity} may not translate to
a measure of distance, i.e., of {\em dissimilarity}, that satisfies
the metric conditions necessary for the application of efficient
metric search structures. 

In such cases a computationally expensive
surrogate distance function may be applied as a culling or {\em
gating} strategy to retrieve a superset of candidate objects to
which a more complex and even more computationally expensive measure
of similarity or dissimilarity will subsequently be applied. In this
context it may be expected that a significant fraction of the dataset
may be retrieved, but the goal is to retrieve them as efficiently as
possible before the ``true'' measure is applied to obtain the desired
set of objects. What is critical to note is that in applications of
this type there is no subsequent use made of the distances calculated 
using the surrogate distance function during the search process, so 
the savings obtained from collection are achieved at no cost.
 
 We note that if distances to retrieved objects are required then it
 may seem that collection cannot be productively applied. However,
 the aggregation of objects in a subtree, followed by distance calculations
 performed as a post-processing step, is significantly more efficient
 than incurring the overhead of a full search traversal of the subtree.
 In addition, the post-calculation of distances with respect to the
 query object is trivially parallelizable, whereas the process of 
 performing distance calculations during the full traversal process 
 is not.

\section{CMT Range Query Tests}

\IEEEPARstart{I}{n} this section we examine CMT performance on range
queries to assess the benefits of cascading\footnote{The tree construction 
algorithm, given in Appendix\,\ref{app:bom}, produces a balanced 
binary tree of height $\lceil\log_2 n\rceil$.}. We do this by
comparing with a variant of the CMT range query algorithm 
in which pruning is limited only to ancestral information up to 
the parent node, {\em CMT-1}, and with a conventional linear-space 
metric search structure, referred to as {\em Baseline}, which does
not maintain any ancestral information and serves to represent
the performance of prior linear-space metric search structures
(e.g., metric trees \cite{Uhlmann1991}, VP trees \cite{Yianilos1993},
and their many variants \cite{sametbook})\footnote{A C++ reference
implementation of all algorithms discussed in this paper, including
 the Baseline search structure and algorithms, is
provided at https://github.com/ngs333/CMT.}.

\begin{figure*}
\centering
\begin{subfigure}{.5\textwidth}
  \centering
  \includegraphics[width=1.0\linewidth]{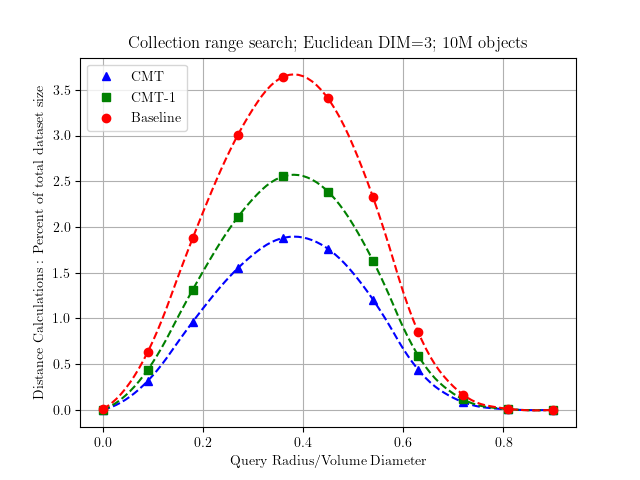}
  \caption{}
  \label{test:rqradius3d}
\end{subfigure}%
\begin{subfigure}{.5\textwidth}
  \centering
  \includegraphics[width=1.0\linewidth]{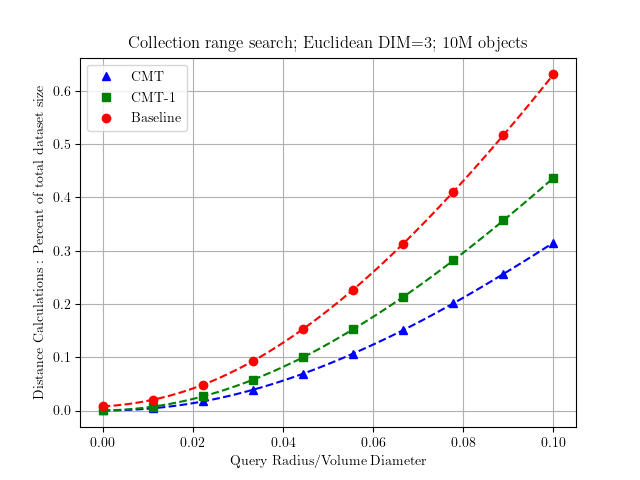}
  \caption{}
  \label{test:rqradius3dz}
\end{subfigure}
\begin{subfigure}{.5\textwidth}
  \centering
  \includegraphics[width=1.0\linewidth]{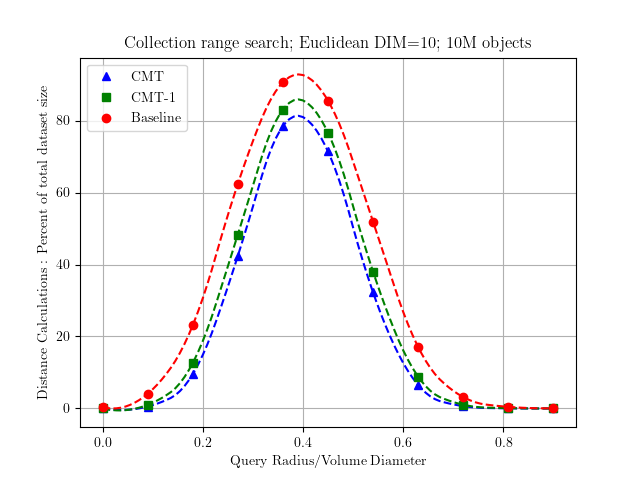}
  \caption{}
  \label{test:rqradius10d}
\end{subfigure}%
\begin{subfigure}{.5\textwidth}
  \centering
  \includegraphics[width=1.0\linewidth]{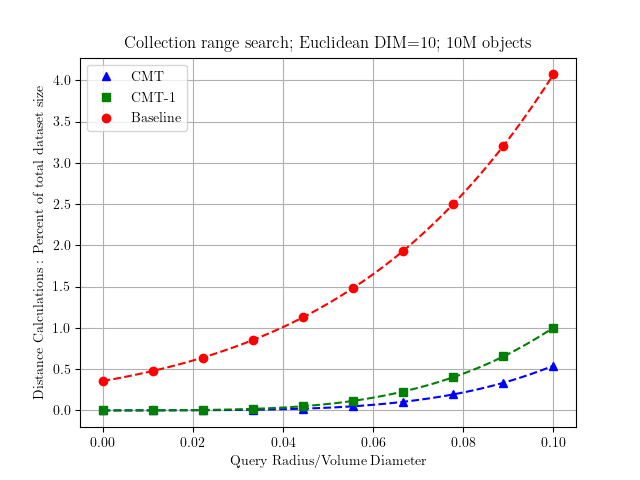}
  \caption{}
  \label{test:rqradius10dz}
\end{subfigure}
\begin{subfigure}{.5\textwidth}
  \centering
  \includegraphics[width=1.0\linewidth]{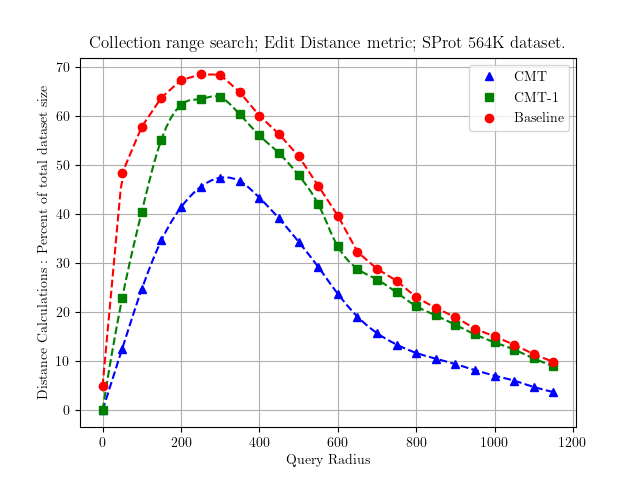}
  \caption{}
  \label{test:rqsprot}
\end{subfigure}%
\begin{subfigure}{.5\textwidth}
  \centering
  \includegraphics[width=1.0\linewidth]{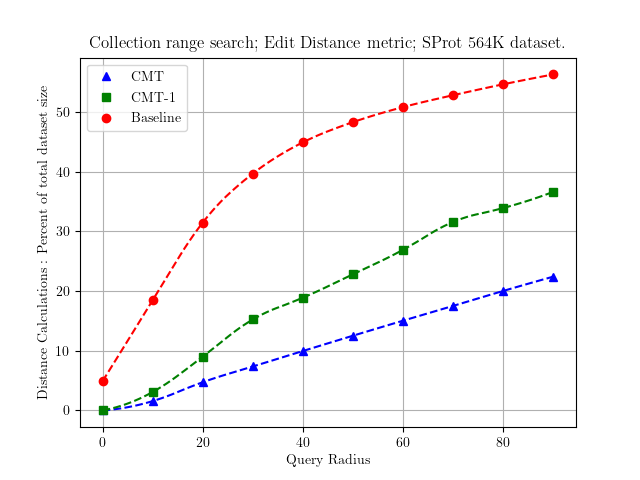}
  \caption{}
  \label{test:rqsprotz}
\end{subfigure}
\caption{{\em Range/Radius Search Performance of CMT for Euclidean and Edit-Distance Metrics} - Each pair of plots shows the relative number of distance calculations needed for each of the compared methods. {\em Baseline} represents conventional linear-space metric search structures, and {\em CMT-1} is CMT restricted to only 1 level of ancestral information. The first plot of each pair shows relative performance as a function of query range/radius from zero until the query volume encompasses the entire dataset. The plateau is where collection/aggregation begins to dominate. The zoomed plots of $(b), (d),$ and $(f)$ show that CMT reduces the number of distance calculations by factors of between $2$ and $5$ over Baseline.}
\label{test:rqs}
\end{figure*}

Our first battery of tests are performed on datasets of $10^7$ uniformly-distributed
Euclidean-space objects in $3$ and $10$ dimensions. Figures \ref{test:rqradius3d} 
and \ref{test:rqradius3dz} show 
the fraction of objects for which distance calculations are performed as a function of
query radius/volume. As the query radius increases, the number of 
distance calculations can be expected to increase until the query volume becomes
so large that collection begins to dominate (i.e., an increasing number of objects
can be identified as satisfying the query without need for individual distance 
calculations) and this is clearly seen. CMT performs roughly half as many
distance calculations as Baseline across most of the range of possible
query radii, and the performance of CMT-1 is roughly betwen that of CMT
and Baseline. The performance of CMT-1 indicates that only one generation
of ancestral information is sufficient to provide more than half of the pruning
power of CMT in this case.

Figures \ref{test:rqradius10d} and \ref{test:rqradius10dz} perform
analogous tests but in $10$ dimensions. Because intrinsic difficulty
of search increases with dimensionality, it is not surprising to see
that the advantage of CMT over Baseline increases from a factor
of $2$ to a factor of $5$ reduction in needed distance calculations.
On the other hand, uniformly-distributed objects in Euclidean space
make the problem relatively easy in the sense that it is amenable
to effective approximate search methods based on grid or orthogonal
range-search structures such as the kd-tree. This may partly explain
why the performance of CMT-1 degrades less relative to that of CMT
when the dimensionality is increased from $3$ to $10$.

Figures \ref{test:rqsprot} and \ref{test:rqsprotz} show results
for edit-distance (more specifically Levenshtein 
distance \cite{10.1093/bioinformatics/btw753,10.1145/375360.375365})
range queries on the 2021 SwissProt database of $564$K protein
sequences \cite{swissprot}, with an average protein length
of 360 amino acids. Figure \ref{test:rqsprot} shows that 
the performance of CMT-1 is degraded significantly relative to CMT 
and approaches that of Baseline. This tends to suggest that
CMT-1's lack of cascaded information beyond the parent 
significantly limits its pruning power, i.e., pruning information 
provided by distant ancestors for CMT tends to increase with 
problem difficulty.

Figures \ref{test:rqradius3d}, \ref{test:rqradius10d}, and
\ref{test:rqsprot} show the power of collection for large
retrieval sets due to large query volumes. Specifically, the
number of distance calculations eventually plateaus and 
then diminishes with increasing search radius. Although 
conventional metric search structures examined in the
literature have not exploited collection capability, we have
equipped the Baseline range search algorithm to exploit
collection. Without it, the number of distance calculations
increases exponentially with query radius and cannot
be meaningfully graphed against searches performed
with collection.

\section{{\em k}-Nearest Neighbor (kNN) Queries}

\IEEEPARstart{T}{he} $k$-nearest neighbor query (or $K(q,k)$) 
asks for the $k$ objects in a dataset $S$ that are nearest to a
given query object $q$. Unlike range queries, for which the
order of tree traversal has no effect on performance (i.e.,
number of required distance calculations and/or node visitations), 
the ordering of
nodes visited for kNN queries has a substantial effect. This 
is because the bounding radius for the currently-identified 
nearest $k$ objects during the search imposes a pruning
constraint on how close future objects must be to $q$ in
order to displace the farthest element of the current set of
$k$ nearest objects. Thus, the sooner a ``good'' set of $k$
objects can be found, the sooner non-candidates can be
pruned.

A priority queue provides precisely the needed means
for steering the sequence of nodes visited to those with
higher likelihood of containing objects nearer to $q$.
This is achieved by computing a ``priority'' for each 
visited node and placing the node (with its priority) into 
the queue. The next node to be visited is then obtained
as the node with highest priority in the queue. The 
resulting performance of the overall algorithm thus
depends, of course, on the formula used to determine 
the priorities.

Given the distance $d(p,q)$ between query object $q$
and node object $p$, and the bounding radius for
for $p$ that contains all objects in its associated
subtree, a triangle inequality calculation can 
produce a lower bound on the distance to the 
nearest object in the subtree. Alternatively, a
lower bound can be obtained for the distance of
the nearest possible object outside the bounding
radius of the node\footnote{CMT nodes actually
contain more precise information in the form of
both the minimal bounding radius for objects within
the subtree {\em and} the maximal radius that
does not include another object in the dataset.
Appendix\,\ref{app:kNNquery} includes details
for how this information can be exploited to achieve 
tighter lower-bound distances to nearest objects 
depending on whether $q$ is within or outside 
of the node's bounding volume.}. 

In the case of non-cascaded metric search trees,
a lower-bound distance between $q$ and the 
nearest-neighbor in the subtree of the current node 
can be directly calculated. It is zero if $q$ is within
the node's bounding volume; otherwise it is the 
distance from $q$ to the maximum bounding 
volume that is guaranteed to not include and
objects within the bounding volume of the
node. In either case, the computed value is the 
obvious choice for use as a search priority.
What is critical to note is that this leads to what is
essentially a proximity-based greedy search of the
tree.

For CMT, by contrast, multiple lower-bound
distances can be computed: one from the current
node, and one from use of the triangle inequality
with the available distances $d(q,a_i)$, $d(a_i,p)$, 
and $d(q,p)$ for each ancestor object $a_i$. Because
each represents a lower bound, the {\em largest}
of the lower bounds represents the most informative
estimate of the smallest distance to an object in
the subtree, and it is used as the priority for the
node by the CMT kNN query algorithm (given in
Appendix\,\ref{app:kNNquery}). In other words,
the priority function exploits information available
in the CMT that is not available to structures of linear
$O(N)$ size. This permits the kNN search algorithm
to pursue a risk-averse ordering of nodes rather 
than a greedy proximity-focused ordering.

\subsection{Range-Optimal kNN Search}

In the computer science literature, many data structures 
for satisfying {\em orthogonal} (also known as 
{\em coordinate-aligned} or {\em box}) queries have
been developed that are amenable to proving
rigorous worst-case complexity bounds on the search time. 
For example, the optimal linear-size range-search data 
structure \cite{bentley1975multidimensional} can 
provably satisfy multidimensional 
range queries in $O(N^{(1-1/d)}+m)$ time, where 
$d$ is the dimensionality of the space and $m$ is the 
number of retrieved objects.

Unfortunately, few rigorous claims can be made
about the performance of metric search structures
without reference to properties of the specific 
metric that will be used. This presents a challenge
to metric search as a research area because 
performance can typically only be expressed in the 
form of limited empirical comparisons to other methods.
We now propose a step forward in this regard 
by defining the optimality of a given kNN algorithm
for a particular metric search structure in terms of 
its performance relative to a range query (without
collection) on that
search structure with assumed knowledge of 
the minimum-size bounding ball containing the 
$k$-nearest neighbors.

\begin{quote}
{\bf Definition}: A $k$-nearest neighbor search algorithm 
is {\em range-optimal} if its empirically assessed 
performance (e.g., in terms of distance calculations
or number of nodes visited) 
approaches that of the best range query algorithm 
for the search structure of interest when given the
smallest search radius enclosing the $k$-nearest 
objects.   
\end{quote}

The rationale for this definition is that
the satisfaction of a range query is not sensitive to the
order in which the satisfying objects are identified during
the search process. More specifically, whether or not a
particular object satisfies the query can be determined
in isolation (independently) without regard to knowledge 
about other objects in the dataset. In other words, the 
sole information available for pruning is given by the 
radius of the range query.

By contrast, a kNN search initially has no pruning
information and therefore must
dynamically {\em acquire} it in the form of a
variable bounding radius at each point during
the search process based on the objects examined
up to that point. This means that an optimally-defined
range query should represent a lower bound on the
best performance possible for satisfying a kNN
query\footnote{Said another way, a kNN algorithm
that outperforms a given range query algorithm
with the optimal $k$-enclosing radius should be
interpreted as evidence {\em that the range query
algorithm is suboptimal}.}.

It should be noted that optimality as defined here
is search-structure-specific in that kNN performance
is bounded by the best possible range-search
algorithm {\em for that structure}. As an example, 
an unstructured dataset of size $n$ offers range-optimality 
in the trivial sense that both kNN and range searches 
must perform exactly $n$ distance calculations during
an exhaustive brute-force examination of all objects
in the dataset.

The value of the range-optimal kNN criterion is realized
when assessing the relative performance of different
priority-search methods for satisfying kNN queries
on sophisticated data structures. Specifically, if a kNN 
method is found that achieves the same performance 
as a minimum-radius range query for the $k$-nearest 
neighbors, no other method can possibly perform
better. 

During the kNN tests discussed in the next section,
we assessed the range-optimality of CMT and Baseline. 
We found CMT to be 99\% range-optimal, which means 
that improvements to the CMT kNN search algorithm can
at most reduce the number of distance calculations and/or
node visitations by a small fraction. Baseline was similarly 
found to be between 97\% and 99\% range-optimal.  
This implies that empirically-observed performance 
advantages of CMT for kNN queries cannot be attributed to 
possible use of a highly suboptimal kNN query algorithm for
the Baseline search tree.

\subsection{Range-Bounded kNN Queries}

As discussed earlier, the query model can be extended to a general
pruning function over any number of application-relevant parameters, e.g.,
\begin{equation}
  f(r,k,nv,dc,t,...) ~<~ \mbox{\em threshold}
\end{equation}
where in isolation a bound $r$ defines an ordinary range query; $k$ 
defines a bound on the number of returned nearest neighbors; $nv$ 
defines a bound on nodes visited; $dc$ defines a bound on distance 
calculations; $t$ defines a bound on total execution time; and so on.
Various multi-criteria pruning methods have been examined, often
termed {\em approximate} queries when the criteria does not
guarantee that the $m$ returned objects are the {\em nearest}
$m$ objects \cite{Mao2006MoBIoSI,4498354}.

Of particular practical interest is the satisfaction of range-bounded
kNN queries of the form $K(b,k)$, where $b$ defines a bound on the
maximum range/radius and $k$ defines the number of nearest neighbors
to be returned from within that bounded range. Its practical significance
derives from the fact that a standard kNN query can be thought of as
a type of dynamic range query in which which the range is progressively
reduced during evaluation of the query based on the bounding radius of 
the current set of $k$-nearest neighbors. This means that the effective
bounding radius is infinite until a first set of $k$ objects has been 
accumulated, and the maximum radius of those initial $k$ objects is likely
to be much larger than that of the final returned set of $k$-nearest 
objects.

In applications for which it is known that objects beyond some radius
are either not of interest, or that the radius is virtually guaranteed to
contain the nearest $k$ objects, then the bound $b$ can provide
substantial pruning capability early in the query process. As will be
demonstrated in the next section, this can lead to 
orders-of-magnitude improvements in performance.


\begin{figure*}
\centering
\begin{subfigure}{.5\textwidth}
  \centering
  \includegraphics[width=1.0\linewidth]{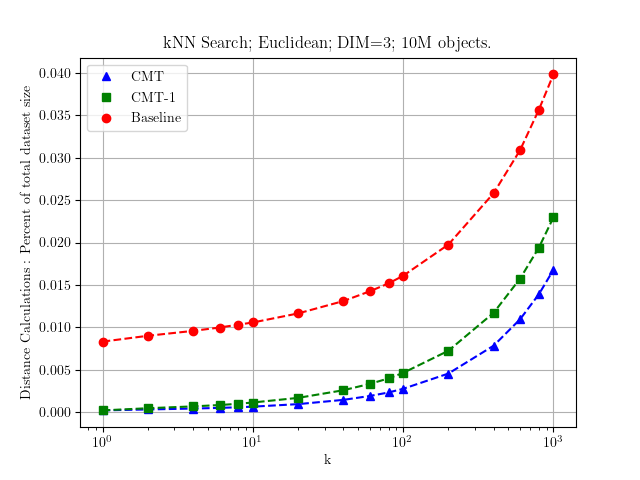}
  \caption{}
  \label{test:knn3d}
\end{subfigure}%
\begin{subfigure}{.5\textwidth}
  \centering
  \includegraphics[width=1.0\linewidth]{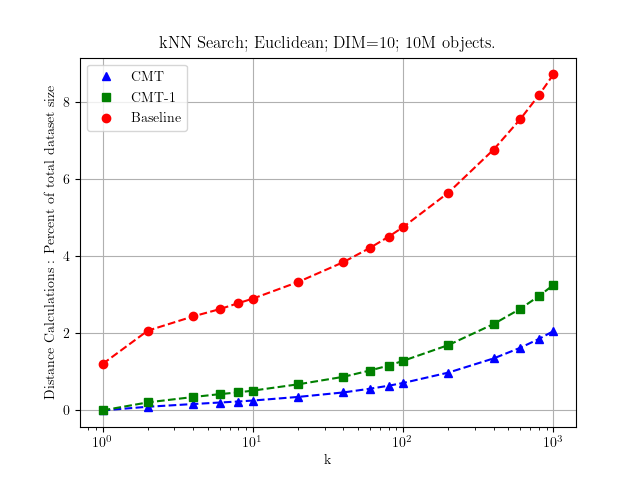}
  \caption{}
  \label{test:knn10d}
\end{subfigure}
\begin{subfigure}{.5\textwidth}
  \centering
  \includegraphics[width=1.0\linewidth]{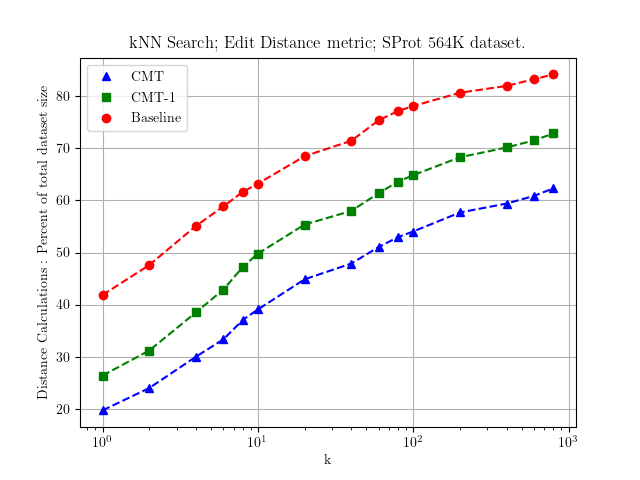}
  \caption{}
  \label{test:knnsprot}
\end{subfigure}%
\begin{subfigure}{.5\textwidth}
  \centering
  \includegraphics[width=1.0\linewidth]{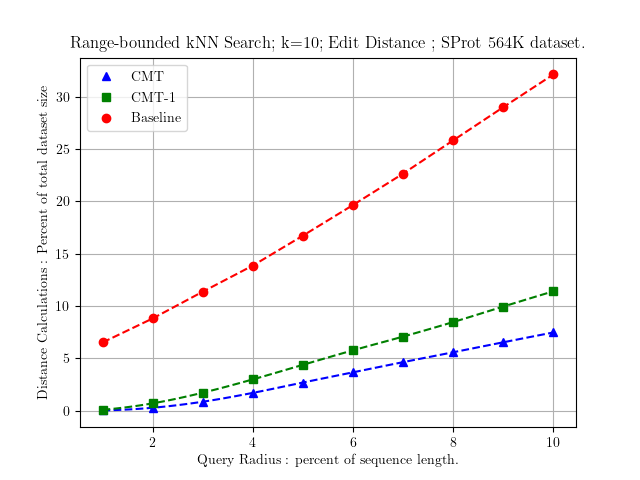}
  \caption{}
  \label{test:sprotseq}
\end{subfigure}%
\caption{{\em kNN Search Performance for Euclidean and Edit Distance Metrics} - Subfigures $(a)$-$(c)$ show the number of distance calculations as a function of the selected number $k$ of returned nearest neighbors, with $(a)$ and $(b)$ showing that Baseline performs a factor of between $4$ and $5$ more distance calculations than CMT in both 3 and 10 dimensions. 
For the for ordinary kNN search with Edit Distance Metric on the SwissProt dataset, CMT queries visit about 20 percent less of the dataset than Baseline (Subfigure ($c$)). For range-bound queries, we see in Subfigure ($d$) that the CMT is a dramatic improvement over Baseline (e.g. for a radius that is two percent of the query length, the improvement is about 30 fold ).}
\label{fig:knnsearch}
\end{figure*}

\section{CMT {\em k}-Nearest Neighbor (kNN) Tests}\label{sec:kNN}

\IEEEPARstart{I}{n} this section we examine the
satisfaction of kNN queries for CMT, CMT-1, and Baseline.
Specifically, Figure\,\ref{fig:knnsearch} shows the
the percentage of the dataset examined (i.e., requiring
distance calculations) as a function of the number $k$ of 
returned nearest neighbors. To more realistically model
practical usage, query objects were not chosen from
the actual dataset. Thus, $k$=$1$ does not necessarily
return an object of distance zero from the query object.

Figure\,\ref{test:knn3d} shows results for Euclidean
kNN queries on a dataset of $10^7$ 
uniformly-distributed points in 3 dimensions for $k$ from 1
to 1000. At $k$=$100$ it can be seen that Baseline
performs over 5 times as many distance calculations
as CMT, and over 3 times as many as CMT-1.
As $k$ approaches 1000, the performance of 
all three methods is degrading rapidly. The relative 
behavior of the three methods in 10 dimensions,
shown in Figure\,\ref{test:knn10d} is qualitatively
similar to that in 3 dimensions, but with almost 
two orders of magnitude more distance calculations
per corresponding value of $k$.

Figure\,\ref{test:knnsprot} shows edit-distance kNN 
results for the same $564K$ protein sequence dataset 
as Figure\,\ref{test:rqs}. However, the discrete
nature of edit distance does not admit a smooth functional
relationship with $k$ because each increment of distance 
leads to an integral increase in the number of sequences
such that there are typically many more than $k$ objects
of strictly-equal distance to the query object,
which accounts for the nonsmooth
graphs in Figure\,\ref{test:knnsprot}. 

Another issue that arises with kNN queries in 
practical applications is that the choice of $k$
may dramatically affect query complexity when
one of the $k$ objects is much farther from the
query object than the others. For example, the 
SwissProt dataset of protein sequences has high 
intrinsic dimensionality due to wide variation
in sequence lengths such that $k$=$10$ can 
require $60\%$ of the dataset to be examined. 
In practice, however, the returning of sequences
that are more than a factor of $2$ larger or smaller
than the query sequence doesn't make sense
because the similarity is essentially equivalent to
that between random sequences.

Figure\,\ref{test:sprotseq} is more practically revealing
by providing results for range-bounded kNN queries, $K(b,k)$,
with $b$ determined as a percentage $p$ of the length of the 
query sequence. For $p$=$10$, the Baseline evaluations are reduced 
from about $60\%$ of the dataset size to about $35\%$, and for 
CMT the reduction is from about $40\%$ to $7\%$. 
For smaller ranges, the advantage of CMT over Baseline 
is even more dramatic. In the case of $p$=$<2$, for example, 
the number of distance calculations performed by CMT is
$30$ times less than that of Baseline.

 
\section{Discussion}\label{sec:discussion}

\IEEEPARstart{I}{n} this paper we have introduced the
notion of cascading as a means for permitting metric 
search algorithms to exploit more pruning information
from distance calculations performed along each 
path of a search tree. The {\em cascaded metric
tree} (CMT) is a concrete realization of this concept, 
and we have provided test results showing the
performance advantages it provides for range (ball)
queries and $k$-nearest neighbor (kNN) queries.

Our tests of the benefits provided strictly by
cascading are novel in that we compare both
against a Baseline metric search tree structure,
which we believe serves as a fair proxy for
prior-art methods,
and a version of CMT, referred to as CMT-1,
with only one level of cascading\footnote{It 
would not have been inaccurate, though
potentially confusing, for Baseline to have 
been referred to as CMT-0 in the sense 
that it does not provide any ancestral
pruning information.}. The resulting test
results give us confidence that independent
comparisons of prior-art methods to the CMT
will corroborate its significant advantages.

A contribution of this paper in the context of
range queries is an emphasis on the importance
of {\em collection} as being not only useful
for reducing distance calculations needed for
retrieval, but also for {\em interactive query
systems}, e.g., that use counting
queries to determine the number of objects
that satisfy a given query without retrieving them. 
Although satisfaction of isolated counting
queries on metric search structures is likely
to be only marginally more efficient than
retrieval queries, the number of counting 
queries performed in interactive data analysis 
applications may be proportionally much
higher because of their iterative-refinement
use in formulating range queries with
practical-sized return sets. 

Another contribution of potentially broader 
interest is our metric-independent 
definition of kNN algorithmic optimality, which we refer 
to as {\em range-optimality}, that can be assessed for
any kNN algorithm as applied with any metric search
structure for any choice of metric distance function.
This is a step toward establishing more rigorous 
and objective statements about the performance
of particular metric search structures and query 
algorithms.

A potentially more important practical contribution
of the paper is our set of range and kNN edit-distance tests on 
the Swiss-Prot-X dataset, which contains 564k protein 
sequences. These tests remain far-removed from capturing
the full scope and detail of the kinds of queries needed for
practical sequence analysis, but they provide strong
evidence of the {\em potential} to significantly outperform
optimized brute-force methods as dataset sizes
continue to increase.  

Future work is needed to more fully assess the
query requirements for large-scale use of metric
search structures for applications relating to
proteomic and genomic sequence analyses.
These applications often require the finding of
objects that are within a specified threshold on
{\em similarity} -- rather than {\em distance} -- 
according to a highly complex non-metric 
similarity function. 

Methods for converting similarity queries of this 
kind to metric distances have been  
examined in the bio-medical literature, but 
they tend to achieve only a relatively loose
correspondence, i.e., the resulting metric search 
query is expected to return a superset of the desired 
objects. If this looseness proves unavoidable, the 
need for highly efficient metric search algorithms 
may be necessary to mitigate the resulting 
computational overhead.

Next-generation genomic and proteomic applications
may also demand use of much more sophisticated
metrics such as tree-edit \cite{ECIR-2013-LaitangPB} 
and graph-edit distance
functions \cite{10.1007/978-3-540-89689-0_33,
VLDB-2009-ZengTWFZ}. 
Tree-edit distance, which determines the
number of primitive edit operations necessary to 
transform one tree structure to another, can be
expected to require $O(N^3)$ rather than the
$O(N^2)$ complexity of string-edit 
distance \cite{10.1145/2746539.2746612,
Andoni2010PolylogarithmicAF,
10.1007/978-3-319-68474-1_11,
Pawlik:2011:RRA:2095686.2095692}.
Graph-edit distance, by contrast, has been 
proven to be NP-Complete 
\cite{abuaisheh:hal-01168816}, and thus can be
expected to demand computation time that
is exponential in the size of the molecular
structures contained in the search structure.
Consequently, approximation methods may be
necessary for both tree and graph-edit distance
calculations \cite{DBLP:series/acvpr/Riesen15}, 
despite the possibility that queries
(e.g., performed using CMT) may not equate
to what would be obtained using the true
distance functions.

In summary, we believe the results in this paper
serve to advance the practical application of 
metric search algorithms to a wider array of
real-world problem domains. Of particular 
interest are the enormous datasets  
presently being generated in biomedical, 
astronomical, and experimental physics 
applications. In such applications, reducing
the number of distance calculations by only
a factor of $4$ or $5$ can reduce the total 
processing time for a large data analysis 
program from a month down to a week.

 
\appendices

\section*{APPENDICES}

Sufficient detail is provided in the main text to understand and 
implement the CMT and its associated search algorithms. 
For completeness, however, we provide the following appendices 
with explicit pseudocode based on our C++ reference
implementation.  

\section{Basic CMT Range Query Algorithm}\label{app:BasicRQ}

Pruning operations based on metric triangle-inequality tests can be 
neatly encapsulated for implementation using the concept of 
{\em pruning distances}. Consider the set $C$ of all objects 
in a node's subtree, and let {\em near} and  {\em far} be the distances 
from the node object to the nearest and farthest objects in $C$.
The pruning distance is the distance the query object is outside 
of interval $D \coloneqq \left[ \mbox{\em near,\, far} \right]$ (see Algorithm\,\ref{alg:prunedist}). 
When a node is visited during the search, if the search radius is less
than the pruning distance, the node's subtree can be excluded from the
search. We extend the concept for use with ancestral distance intervals 
and define {\em maximum pruning distance} as  the maximum of the pruning
distances over the ancestral distance intervals.

\begin{algorithm}
	\SetKwFunction{PruningDistance}{\textbf{PruningDistance}}
	\SetKwData{Node}{node}
	\SetKwData{Near}{near}
	\SetKwData{Far}{far}
	\SetKwData{DAQ}{distancePQ}
	\SetKwData{pruningDistance}{pruningDistance}
	\SetKwBlock{Begin}{begin}{end}
		
	\PruningDistance{\DAQ , \Node, l} \Begin(){
	
	\If {$\DAQ < \Node.\Near[l]$} {
		\pruningDistance $\leftarrow \Node.\Near[l] - \DAQ$
	}
	\ElseIf {$\DAQ > \Node.\Far[l]$}{
		\pruningDistance $\leftarrow  \DAQ - \Node.\Far[l]$
	}
	\Else {
		\pruningDistance $\leftarrow  0$
	}
	\KwRet{\pruningDistance}
	}
	\caption{\Vspcx Calculation of the pruning distance by the function 
	            $PruningDistance( d_{pq}, n, 0)$.}
	\label{alg:prunedist}
\end{algorithm}

Algorithm \ref{alg:BasicRangeQuery} performs basic CMT range queries.
Its recursive structure applied to the tree visits a node at most once.
Upon visiting a node, the max pruning distance with respect to the
set of ancestral objects is determined, if this is greater than the
query radius then the current search path can be terminated.
If the query object is not pruned at this step, then a distance
calculation must be performed between the query object and the
node object, which provides one more pruning test with respect
to the bounding volume of the node. If the node object satisfies the 
query then it is added to the result set. Regardless, the distance
between the query object and the node object is added to the list
of ancestral objects for use in pruning tests at subsequently visited
nodes along the current path in the tree, and the search
proceeds recursively to the left subtree and then the right
subtree.

  \begin{algorithm}
    \SetKwFunction{BasicRangeQuery}{\textbf{BasicRangeQuery}}
    \SetKwData{Object}{object}
    \SetKwData{MaxPDF}{MaxPruningDistance}
    \SetKwData{MaxPDV}{maxPD}
    \SetKwData{PDF}{PruningDistance}
    \SetKwData{PD}{PD}
    \SetKwData{Stack}{stack}
    \SetKwData{SQ}{query}
    \SetKwData{Dist}{distance}
    \SetKwData{MDI}{MDI}
    \SetKwData{Node}{node}
    \SetKwData{Left}{left}
    \SetKwData{Right}{right}
    \SetKwData{null}{null}
    \SetKwBlock{Begin}{begin}{end}
    \BasicRangeQuery{\SQ,  \Node, \Stack} \Begin(){
	
	\If { \Node is \null}{
	  return 
	}
	
	\MaxPDV $\leftarrow$ \MaxPDF(\Stack, \Node)	
	
	\If { $\MaxPDV \leq \SQ.searchRadius()$ }  {
	  
	  \Dist $\leftarrow  \Dist(\SQ.\Object, \Node.\Object) $
	  
	  \If {$\Dist \leq \SQ.searchRadius()$ }{
	    \SQ.addResult(\Node.\Object, \Dist)
	  }
	  
	  \PD = \PDF(\Dist, \Node, 0)
	  
	  \If { $\PD \leq \SQ.searchRadius()$ }{
	    
	    \Stack.push( \Dist )
	    
	    \BasicRangeQuery(\SQ, \Node.\Left, \Stack )
	    
	    \BasicRangeQuery(\SQ, \Node.\Right, \Stack )
	    
	    \Stack.pop()
	  }
	}
	\KwRet{}
      }
      \caption{Basic CMT Range Query Algorithm}
      \label{alg:BasicRangeQuery}
  \end{algorithm}


\section{Collection CMT Range Query Algorithm}\label{app:collectRQ}

{\em Collection} is an enhancement to the
basic range query (Algorithm \ref{alg:BasicRangeQuery}) with
tests added to identify cases in which the query volume encloses a node's 
bounding volume. We define the {\em collection distance} as the sum of 
the distance between the query object and the node object, plus the {\em far} 
component of the node's distance interval. We also define a 
{\em minimum collection distance} as the minimum of the collection 
distances over the ancestral distance intervals. More intuitively, this
represents the intersection of constraining bounding volumes of the
ancestor objects and that of the current node. If the query volume
contains this intersection volume then all objects in the subtree
rooted at the current node must necessarily satisfy the query, and
therefore they may be collected into the result set without need for
explicit distance calculations.

Algorithm \ref{alg:CollectRangeQuery} performs CMT queries with collection.
The algorithm is similar to algorithm \ref{alg:BasicRangeQuery} except for
the addition of a test against the minimum collection distance and a test against 
the collection distance. When the distances are less than the search radius, 
all objects in the subtree (including the node object) are added to the result set. 

\begin{algorithm}
  \SetKwFunction{CollectRangeQuery}{\textbf{CollectRangeQuery}}
  \SetKwData{Object}{object}
  \SetKwData{MaxPDF}{MaxPruningDistance}
  \SetKwData{MaxPDV}{maxPrunigDist}
  \SetKwData{MinCDF}{MinCollectionDistance}
  \SetKwData{CDF}{CollectionDistance}
  \SetKwData{MinCDV}{minCD}
  \SetKwData{CDV}{cd}
  \SetKwData{PDF}{PruningDistance}
  \SetKwData{PD}{pruningDist}
  \SetKwData{Stack}{stack}
  \SetKwData{SQ}{query}
  \SetKwData{Dist}{distance}
  \SetKwData{MDI}{MDI}
  \SetKwData{Node}{node}
  \SetKwData{Left}{left}
  \SetKwData{Right}{right}
  \SetKwData{null}{null}
  \SetKwBlock{Begin}{begin}{end}
  \CollectRangeQuery{\SQ,  \Node, \Stack} \Begin(){
      
      \If { \Node is \null}{
	return 
      }
      
      \MaxPDV $\leftarrow$ \MaxPDF(\Stack, \Node)	
      
      \If { $\MaxPDV \leq \SQ.searchRadius()$ }  {
	
	\MinCDV $\leftarrow$ \MinCDF(\Stack, \Node)	
	
	\If { $\MinCDV \leq \SQ.searchRadius()$ }  {
	  \SQ.addSubtree(\Node, \MinCDV)
	}
	\Else{
	  
	  \Dist $\leftarrow  \Dist(\SQ.\Object, \Node.\Object) $
	  
	  \CDV $\leftarrow \CDF(\Dist, \Node) $
	  
	  \If {$\CDV \leq \SQ.searchRadius()$ }{
	    \SQ.addSubtree(\Node, \CDV)
	  }
	  \Else{
	    
	    \If {$\Dist \leq \SQ.searchRadius()$ }{
	      \SQ.addResult(\Node.\Object, \Dist)
	    }
	    
	    \PD $\leftarrow$ \PDF(\Dist, \Node, 0)
	    
	    \If { $\PD \leq \SQ.searchRadius()$ }{
	      
	      \Stack.push( \Dist )
	      
	      \CollectRangeQuery(\SQ, \Node.\Left, \Stack )
	      
	      \CollectRangeQuery(\SQ, \Node.\Right, \Stack )
	      
	      \Stack.pop()
	    }
	  }
	}
      }
      \KwRet{}
    }
    \caption{Full range query algorithm with collection/aggregation.}
    \label{alg:CollectRangeQuery}
\end{algorithm}


\section{CMT {\em k}-Nearest Neighbor (kNN) Algorithm}\label{app:kNNquery}

The CMT kNN query algorithm \ref{alg:kNNQuery} uses a priority 
queue to determine the node visitation order. The
priority queue provides the standard push() and pop()
functions with members given in Table\,\ref{tab:pqnode}.

\begin{table}[h]

\centering
\caption{Data Members of Priority-Queue Node}
\begin{tabular}{r|l}
	\hline
	\textbf{tree node}\Vspcx & The \emph{tree} node associated with the queue node. \\
	\textbf{parent} & The parent queue node of the queue node. \\
	\textbf{distance} & The distance between the tree node object and\\  
	~ & the query object. \\
	\textbf{priority} & Measure of likelihood that close neighbors\\
	~ & will be found in the subtree of the tree node.\\
	\hline
\end{tabular}
\label{tab:pqnode}
\end{table}

The max pruning distance between the tree node and the query object
is used for the priority, causing the search algorithm to preferentially 
visit nodes with greater likelihood of being pruned\footnote{This
is typically done using the distance between the query object 
and the node object as a priority \cite{uhlmannImp1991}.}.

The algorithm visits a tree node by removing a queue
node from the priority queue. The tree node is pruned if its
priority (i.e., max pruning distance) is greater than the 
search radius. The distance between the query and node
objects is then calculated; and if the object is within the
current range then it is added to the result set, and if the
size of the set exceeds $k$ then the farthest of the $k+1$
objects is removed and the minimum bounding radius of the 
remaining $k$ objects is updated. This bounding radius (distance) 
is also used to determine the max pruning distances of the 
two child nodes, and if they cannot be immediately pruned
then they are each placed into the priority queue with their
respective max pruning distances as their priorities. This
means that a pruning test is applied both before the object
is added to the queue {\em and} when it is later removed
from the queue, where the latter test can be productive
if the bounding radius of the $k$ objects has decreased in
the interim.

\begin{algorithm}[h]
  \SetKwFunction{kNNQuery}{\textbf{kNNQuery}}
  \SetKwData{Object}{object}
  \SetKwData{MaxPDF}{MaxPruningDistance}
  \SetKwData{MaxPDV}{maxPruningDist}
  \SetKwData{PDF}{PruningDistance}
  \SetKwData{PQNode}{queueNode}
  \SetKwData{PQNodeType}{QueueNode}
  \SetKwData{PQ}{queue}
  \SetKwData{SQ}{query}
  \SetKwData{Dist}{distance}
  \SetKwData{MDI}{MDI}
  \SetKwData{Node}{node}
  \SetKwData{Left}{left}
  \SetKwData{Right}{right}
  \SetKwData{Priority}{priority}
  \SetKwData{null}{null}
  \SetKwBlock{Begin}{begin}{end}
  
  \kNNQuery{\SQ,  \PQ} \Begin(){
      
      \While {\PQ \textbf{is not empty}} {
	\PQNode $\leftarrow$ \PQ.pop
	
	\Node $\leftarrow \PQNode.\Node$
	
	\MaxPDV $\leftarrow \PQNode.priority $
	
	\If { $\MaxPDV \leq \SQ.searchRadius()$ }  {
	  
	  \Dist $\leftarrow  \Dist(\SQ.\Object, \Node.\Object) $
	  
	  \If {$\Dist \leq \SQ.searchRadius()$ }{
	    
	    \SQ.addResult(\Node.\Object, \Dist)
	    
	  }
	  
	  \PQNode.dist $\leftarrow$ \Dist
	  
	  \If { $\PDF(\Dist, \Node, 0) \leq \SQ.searchRadius()$ }{
	    
	    \If { $\Node.\Left \neq \null$ } {
	      
	      \Priority $\leftarrow$ \MaxPDF(\PQNode, \Node.\Left)
	      
	      \If { $\Priority \leq \SQ.searchRadius()$ } {
		
		\PQ.push(new \PQNodeType (\Node.\Left, \PQNode , \Priority ))
	      }
	    }
	    
	    \If { $\Node.\Right \neq \null $ } {
	      
	      \Priority $\leftarrow$ \MaxPDF(\PQNode, \Node.\Right)
	      
	      \If { $\Priority \leq \SQ.searchRadius()$ } {
		
		\PQ.push(new \PQNodeType (\Node.right, \PQNode , \Priority ))
	      }
	    }
	    
	  }
	}
      }
      
      \KwRet{}
    }
    \caption{CMT $k$NN Query algorithm}
    \label{alg:kNNQuery}
\end{algorithm}


\section{Tree Construction}\label{app:bom}

In the context of spatial search trees, the term ``pivot'' is commonly used to refer to an object used to discriminate between partitions of a dataset. Typically, the pivot object at a node is selected/determined during construction of the tree to define a partition of the dataset, and that object is stored at a node for subsequent use by a search algorithm to prune the traversal of certain subtrees. In the more general case, the pivot objects used for partitioning during the construction process need not be the same objects as those stored for use during the search process. In other words, a distinction must be made between {\em partition} pivots and {\em search} pivots, and we will explicitly distinguish between the two when there is potential ambiguity based on the context of the discussion.

It should be noted that the general the CMT definition does not specifically prescribe the manner of partitioning and selection of search pivots. Ideally, the search performance properties of the tree should be relatively invariant to discretion applied during the construction process, i.e., search efficiency should {\em not} depend on meticulous tailoring of the construction algorithm to exploit assumed characteristics of datasets arising in a particular application. However, the degree to which this ideal is achieved must be empirically assessed by examining different construction algorithms. For our tests we simply choose each partitioning pivot at random and then compute the median distance to
the remaining objects to obtain a balanced partition. If the median distance is not unique, then each object with distance equal to the median will be arbitrarily assigned to one of the two subsets such that equal cardinality is maintained\footnote{In the reference implementation, this functionality is provided by the C++ std::nth-element function.}. This method ensures that the resulting tree is balanced with $O(\log N)$ height, which minimizes CMT memory requirements by limiting the maximum number of query pivots at any given node to $O(\log N)$. We refer to this as {\em balanced object median} (BOM) partitioning.

The CMT construction of  Algorithm\,\ref{algorithm:buildcmt} is preceded by a pre-processing of a set $O$ of objects by allocating a set $S$ of nodes with  $|S| = |O| = N$, with the assigning of one object to each node. The tree is then constructed by recursively applying a procedure that selects a node $p$ from $S$, which will also serve as a root or parent of the current subtree. This is followed by partitioning the set $\{S \setminus \{p\}\}$ into two subsets, $S_l$ and $S_r$, which will be recursively constructed as the left and right subtrees, respectively. In the process of partitioning, the metric distances between $p$ and each object in $\{S_l \cup S_r\}$ are calculated. Additionally, the minimum and maximum of those distances are stored in the {\em near} and {\em far} data members of the node's distance interval. Finally, a vector $p.pDistance{[}{]}$ is stored as a member of each node and contains the distance between the node object and each of the node's ancestor objects. The function ComputeADIV() uses these vectors to compute, for the local root, a distance interval per each ancestor\footnote{ $p.pDistance{[}{]}$ contains information that is only used in the tree construction process and can be deleted after construction, i.e., it is not a necessary member of CMT nodes.}.

\begin{algorithm}
	\SetKwFunction{BuildCMT}{\textbf{BuildCMT}}
	\SetKwFunction{ComputeADIV}{\textbf{ComputeADIV}}
	
	\SetKwData{Cnodes}{cnodes}
	\SetKwData{Pnode}{pnode}
	\SetKwData{Depth}{depth}
	\SetKwData{Dist}{distance}
	\SetKwData{Nearest}{nearest}
	\SetKwData{Furthest}{furthest}
	\SetKwData{DI}{DI}
	\SetKwData{PDL}{pdl}
	\SetKwData{Intervals}{intervals}
	\SetKwData{PD}{pDistances}

	\SetKwData{Root}{root}
	\SetKwData{Left}{left}
	\SetKwData{Right}{right}
	\SetKwData{Null}{null}
	\SetKwData{Nodes}{nodes}
	\SetKwData{Node}{node}
	\SetKwData{Object}{object}
	\SetKwData{DMR}{dmr}
    \SetKwData{Size}{size}
    \SetKwInOut{Input}{input}

	\SetKwBlock{Begin}{begin}{end}
	
	\ComputeADIV{\Pnode, \Cnodes, \Depth } \Begin(){
		
	\For{$l\in \left[0, \Depth \right)$}{
	
	\PDL $\leftarrow \Depth - l -1 $
	
		\Nearest  $\leftarrow \underset{node \in {\Cnodes \cup \Pnode}} {min \{node.\PD[\PDL] \}}$ 
		
	  	\Furthest $\leftarrow \underset{node \in {\Cnodes \cup \Pnode}} {max \{node.\PD[\PDL] \}}$
		
		\Pnode.near[$l$]  $\leftarrow \Nearest $
	
		\Pnode.far[$l$] $\leftarrow \Furthest $
		}
	}
	\caption{\Vspcx Calculating the ancestral distance intervals for node $pnode$  }
	\label{algorithm:computedadiv}

	\BuildCMT{\Nodes, \Depth} \Begin(){
	\If{ $|\Nodes | == 0 $}{  \KwRet{\Null}  }
		
	\If{$ | \Nodes | == 1 $} { 
	    \Root $\leftarrow$ first \{ \Nodes \} 
	    
	    \KwRet{ \Root }
	    }
        
	\Root $\leftarrow$ a random node $\in$ \Nodes

    \Nodes $\leftarrow \{ \Nodes \setminus \Root \} $
    
	
	\For{\Node $\in$ \Nodes}{				
		\Node.\Dist[ \Depth ] $\leftarrow \Dist ( \Node.object, \Root.object ) $
	}
	
	\Root.near[0]  $\leftarrow \underset{\Node \in \Nodes }{\mathrm{min}\{ \Node.\Dist[ \Depth ] \}}$
		
	\Root.far[0]  $\leftarrow \underset{\Node \in \Nodes }{\mathrm{max}\{ \Node.\Dist[ \Depth ]  \}}$
	
	MedDist $\leftarrow$ {\em median distance to random pivot}
		
	\Left $\leftarrow$ \emph{\{$\Node \in \Nodes \mid \Node.\Dist[ \Depth ] < $ MedDist \}}
			
	\Right $\leftarrow \{ \Nodes \setminus \Left  \} $

	\Root.right$\leftarrow$\BuildCMT{\Right, $\Depth + 1$}
	
	\Root.left$\leftarrow$\BuildCMT{\Left, $\Depth + 1$}

    ComputeADIV ( \Root, \Nodes, \Depth)
		
	\KwRet{\Root}
	
	}
	\caption{\Vspcx ComputeADIV() - Building the CMT with random pivots and BOM partitioning.}
	\label{algorithm:buildcmt}
\end{algorithm}

ComputeADIV() (Algorithm\,\ref{algorithm:computedadiv}) computes ancestral distance intervals for the current node. This algorithm is executed for a given node after its children are recursively processed by the calling algorithm. When this algorithm is called for node $pnode$ at depth $d$ and with descendant node set $cnodes$, each descendant node already has $d+1$ distance values in its own $pDistances$ array. (The index into $pDistances$ starts with the root node of the entire tree indexed as $nd.pDistance[0]$.) The construction algorithm for the {\em Baseline} tree is nearly identical to Algorithm\,\ref{algorithm:buildcmt}, and the number of distance function evaluations is the same, but no ancestral information is maintained. 

\subsection{Tree construction complexity} \label{sec:buildcomp}

The computational complexity of building CMT with BOM
partitioning is technically $O(N \log^2 N)$ because the
tree has $O(N)$ nodes with $O(\log N)$ scalar ancestral
distance values. However, the $O(N \log^2 N)$ component
of the runtime is simply due to the maintenance of
$O(\log N)$ ancestral distances per node. Thus the
coefficient of the $O(N \log^2 N)$ component is relatively
small compared to that of the $O(N\log N)$ distance calculations
performed by ComputeADIV(). Consequently, for any nontrivial distance
function the practical build time for the tree 
will be dominated by the $O(N \log N)$ component.

 \ifCLASSOPTIONcompsoc
  \section*{Acknowledgments}
\else
  \section*{Acknowledgment}
\fi

Thanks to Seth Wiesman and Yeshwanthi Pachalla for their contributions to earlier implementations and tests of the data structure. A portion of this work was sponsored by the Army Research Laboratory and was accomplished under Cooperative Agreement Number W911NF-18-2-0285. The views and conclusions contained in this document are those of the authors and should not be interpreted as representing the official policies, either expressed or implied, of the Army Research Laboratory or the U.S. Government. The U.S. Government is authorized to reproduce and distribute reprints for Government purposes notwithstanding any copyright notation herein.

\bibliographystyle{IEEEtran}

\bibliography{bibtex_entries}

\end{document}